\documentstyle[12pt,diagram]{article}
\makeatletter

\newbox\unobox
\setbox\unobox=\hbox{\kern+.3em {\rm R}%
\kern-0.88em {\rm I\kern+0.9em}}
\def\erreuno{\copy\unobox}
\newbox\duebox
\setbox\duebox=\hbox{\kern+.3em {\rm R}%
\kern-0.88em {\rm I}%
\kern+.5em\raise.88ex\hbox{\scriptsize 2}\kern+.3em}
\def\erredue{\copy\duebox}
\newbox\trebox
\setbox\trebox=\hbox{\kern+.3em {\rm R}%
\kern-0.88em {\rm I}%
\kern+.5em\raise.88ex\hbox{\scriptsize 3}\kern+.3em}
\def\erretre{\copy\trebox}
\newbox\quattrobox
\setbox\quattrobox=\hbox{\kern+.1em {\rm C}%
\kern-0.44em\raise.025em\hbox{\vrule width 0.03 em height 0.65 em}\kern+.5em}
\def\compl{\copy\quattrobox}
\newbox\cinquebox
\setbox\cinquebox=\hbox{\kern+.3em {\rm R}%
\kern-0.88em {\rm I\kern+0.52em}}

\newbox\seibox
\setbox\seibox=\hbox{\kern+.3em {\rm N}%
\kern-0.88em {\rm I\kern+0.9em}\kern-0.4em}
\def\Enne{\copy\seibox}
\newbox\settebox
\setbox\settebox=\hbox{\kern+.1em {$\scriptstyle {\rm C}$}
\kern-0.64em\raise.021em\hbox{\vrule width 0.03 em  height 0.42
em}\kern+.5em}

\newbox\ottobox
\setbox\ottobox=\hbox{\kern+.2em {$\scriptstyle {\rm R}$}%
\kern-0.66em {{$\scriptstyle {\rm I}$}\kern+0.52em}}
\def\errepic{\copy\ottobox}
\def\tfract#1/#2{{\textstyle{\raise0.8pt\hbox{$\scriptstyle#1$}\over%
\hbox{\lower0.8pt\hbox{$\scriptstyle#2$}}}}}
\def\mezzo{\tfract 1/2 }

\begin{document}

\begin{titlepage}
\begin{flushright}
 IFUP-TH. 28/98\\
  July  1998\\
\end{flushright}

\vspace{8mm}

\begin{center}

{\Large \bf Classical Teichm\"uller Theory

\vskip 0.5truecm

and (2+1) gravity}

\vskip 0.7 truecm

\vspace{12mm}
{\large \bf Riccardo  Benedetti ${}^{a}$ ~and~  Enore   Guadagnini
${}^{b}$}

\vskip 0.4 truecm

\vspace{6mm}
($a$) Dipartimento di Matematica dell'Universit\`a di Pisa \\
Via F. Buonarroti, 2. $\; $ I-56100 PISA. Italy

\vspace{3mm}
($b$) Dipartimento di Fisica dell'Universit\`a di Pisa \\
Via F. Buonarroti, 2. $\; $ I-56100 PISA. Italy

\vspace{6mm}
E-Mail: benedett@DM.UNIPI.it ~,  $\;$ guadagni@mailbox.difi.unipi.it

\vspace{3mm}

\end{center}
\vspace{10mm}

\begin{quote}
\hspace*{5mm} {\bf Abstract.}  ~We consider classical Teichm\"uller
theory and  the geodesic flow on  the cotangent bundle of the
Teichm\"uller space. We show that the corresponding orbits provide a
canonical description of  certain (2+1) gravity systems in which a set
of point-like particles evolve in universes with topology $ \Sigma_g
\times \erreuno   $ where $ \, \Sigma_g  $ is a Riemann  surface of
genus $ \, g >1\, $.   We construct an explicit York's slicing
presentation  of the associated spacetimes, we give an interpretation of
the asymptotic states in terms of measured foliations and discuss the
structure of the phase spaces.

\end{quote}
\end{titlepage}
\clearpage

\noindent {\bf 1. Introduction.} ~Gravity in $\, (2+1)\, $ dimensions
can be  understood as a  non trivial  toy model of the  physically
significant $(3+1)-$dimensional case.   Einstein equations of $\, (2+1)\,
$ gravity  simplify because the Ricci tensor determines the Riemann
tensor: this $(2+1)-$dimensional property  allows also for an  approach
to the problem which is much more geometric than analytic. A matter-free
solution of topological type  $\, \Sigma_g \times \erreuno  $,  where
$\, \Sigma_g \, $ is a compact surface of genus $\, g\, $, will be
called an empty universe  of type $\, g\, $ and  is a locally
Minkowskian 3-manifold fibred by the spatial surfaces $\,
\Sigma (t) = \Sigma_g \times \{ t \}\, $.

Two main approaches have been used to study empty universes of type  $
\, g \, $:  one \cite{1,2} is based on the holonomy representations of $
\, \pi_1 (\Sigma_g ) \, $ in the Poincar\'e group $ \, ISO(2,1) \, $ and
the other \cite{3,4} makes use of the canonical ADM construction. Both of
them identify the cotangent bundle $ \, T^*_g \, $ of the Teichm\"uller
space $\, T_g \, $ as the phase space of the empty universes of type $
\, g \, $. In fact, the precise correspondence between these two
identifications is explicitly known \cite{5} only for genus $ \, g =1 \,
$; for $ \, g >1 \, $ several open questions remain to be answered.

In this article we shall show that classical Teichm\"uller theory also
allows  for an interpretation of $ \, T^*_g \, $ as a subset of the
phase space of certain (2+1) gravity systems with ``matter".  More
precisely we will show that, with respect to the dynamics described by
the Teichm\"uller geodesic flow in $\, T^*_g \, $ for $ \, g >1 \, $,
each Teichm\"uller line gives a canonical description of the time
evolution of a particular set of point-like gravitating  particles
moving on  a 2-dimensional surface of genus $ \, g \,
$.  In the  $ \, g=1 \, $ case, our construction reproduces the known
\cite{2,4,5} matter-free solutions.

Let us give a short description of the main geometric idea. Our result
is based on the  properties of the Teichm\"uller deformations of the
conformal structure associated with a Riemann surface. For each
Teichm\"uller line, the corresponding deformations are the result of a
stretching of the leaves of a particular measured foliation. This
foliation determines a canonical flat structure on the surface with
conical singularities which can be interpreted as point-like particles.
The effect of these particles  is to ``flatten" spatial slices. The
parameter which determines the strength of the stretching is identified
with ``time" and, with the spatially flat metric on each time slice, one
can  construct a spacetime which (in the complement of the particle
world-lines) is flat. Moreover, the time slices have constant extrinsic
curvature (again, far from the singularities) which is simply related to
the  time  parameter; so, our construction provides a natural York
slicing of spacetime.

In the first part of this article, we summarize a few basic results of
classical Teichm\"uller theory and the geometry of quadratic
differentials \cite{6,7,8,9} for Riemann surfaces of genus $  g \geq 1 \,
$. Then, we construct the spacetimes of the  gravity systems associated
with Teichm\"uller lines and we produce  an explicit York slicing
description of each of them.  Finally, we give an interpretation of the
associated asymptotic states in terms of measured foliations on $
\Sigma_g  $ and we discuss the structure of the phase space.

\medskip

\noindent{\bf 2. The geometry of quadratic differentials and
Teichm\"uller theory.}  ~Let $ \, S \, $ be a compact Riemann surface of
genus $ \, g \geq 1 \, $. A  quadratic differential $ \, \omega \,
$ on $ \, S \, $ is a holomorphic map  $ \, \omega \, : \, TS \to \compl
$, where $ \, TS \, $ is the complex tangent bundle on $ \, S \, $, such
that for any point $ \, p \in S \, $ the restriction of $ \, \omega \, $
to $ \, TS_p \, $ is a quadratic form. In a system of local complex
coordinates $ \, z = x + i y \, $, we write $ \, \omega = \varphi (z) \,
dz^2 \, $.  Let us denote by $ \, Q(S) \, $ the set of quadratic
differentials on $ \, S \, $; $ \, Q(S) \, $ is a complex vector space
and, by Riemann-Roch Theorem, it has complex dimension  $\,   3 \, g \, -
\, 3  \, $  when $ \, g \geq 2 \,
$ and complex dimension 1 for $ \, g =1 \, $.  One can introduce a norm
on $ \, Q(S)\, $ according to
$$
\| \omega \| \; = \; \int_S \, | \omega | \; = \; \int_S \, | \varphi |\,
dx \,  dy \qquad .
$$

Any given quadratic differential $ \, \omega \in Q(S) \backslash \{ 0 \}
\, $ selects a particular set of points on $ \, S \, $ in which $ \,
\omega \, $ vanishes.  If $ \, p \in S \, $ is a zero of
$ \, \omega \, $, one can find a local coordinate $ \, z \, $ at  $ \, p
\, $, with $ \, z(p) = 0 \, $, such that
$$
\omega \, = \, z^m \, dz^2  \qquad .
$$
The positive integer $ \, m\, $ is the  order of $ \, \omega \, $ at  $
\, p \, $ and the coordinate $ \, z \, $ is unique up to rotations of
angles $ \, 2 \pi n / (m+2) \, $ with integer $ \, n \, $.

Let $ \, \omega \in Q(S) \backslash \{ 0 \} \, $; a vector $ \, X \in
TS_p \, $  is  called $ \, \omega$-horizontal if  $ \, \omega (X) \, $ is
real and strictly positive.  The vector  $ \, Y \in TS_p \, $  is  $ \,
\omega$-vertical if  $ \, \omega (Y) \, $ is real and strictly negative.
Let us now consider the set $ \, {\widehat S } = S \, \backslash \, \{
\,  {\rm zeros ~ of ~ } \omega \} \, $. For every $ \, p \in {\widehat S }
\, $, $ \, TS_p \, $ has one $ \, \omega$-horizontal and one  $
\, \omega$-vertical directions which are mutually orthogonal with respect
to the Riemannian  $\, \omega$-metric on $ \, {\widehat S } \, $
$$
d s^2_\omega \; = \; |\, \varphi \, | \, | \, d \, z \, |^2 \qquad .
$$
The integral lines of these two fields of directions are the leaves of
the $ \, \omega$-horizontal and $ \, \omega$-vertical foliations on $ \,
{\widehat S} \, $. Both the $ \, \omega$-foliations and the $ \,
\omega$-metric extend to the whole surface $ \, S \, $ with
singularities  at the zeros of the quadratic differential $ \, \omega \,
$.  We shall now describe the geometry induced on $ \, S
\, $ by the  $ \, \omega$-metric and the associated $ \,
\omega$-foliations. We shall firstly  concentrate on $ \, {\widehat S}
\,  $ and then we shall analyze the singular points.

In a neighbourhood of any $ \, p \in {\widehat S}\, $ where the
coordinate  $ \, z \, $   satisfies $ \, z(p) = 0 \, $, consider the
change of coordinates
$$
\zeta (z) \; = \; \int^z \sqrt \omega \; = \; \int_0^z \, \sqrt
{\varphi (z^\prime)} \, dz^\prime
\qquad ,
$$
where a fixed choice of the sign for the square root has been made.  If $
\, \zeta = \chi + i \eta \, $, the $ \, \omega$-horizontal lines become $
\, \{ \eta = {\rm constant } \, \} \, $ and the $ \, \omega$-vertical
lines correspond to $ \, \{ \, \chi = {\rm constant} \, \} \, $.
Moreover, $ \, \omega = d \, \zeta^2 \, $ and $ \, ds^2_\omega = d\chi^2
+ d \eta^2 \, $. This makes clear that the Riemannian $ \, \omega$-metric
on $ \, {\widehat S} \, $ is flat.

Assume now that $ \, p \, $ is a zero of $ \, \omega \, $ and let $
\, \omega = \, z^m \, dz^2 \, $ in a chart such that $ \, z (p) =0 \, $.
Let us distinguish two possibilities.  When $\, m\, $ is even, $ \,
\sqrt \omega \, $ has a single valued branch and $ \, \zeta = \zeta (z) \,
$ is a $\, (m+2) /2  $-sheeted branched covering ramified over 0.
Consequently, as far as the horizontal lines are concerned, $ \,
\zeta^{-1} (\erreuno ) \, $ consists of $ \, ( m+2 )\, $ analytic rays
emanating from  0 which are equally spaced. One finds a similar
structure  for the  vertical lines. When $\, m \, $ is odd, one can
reduce the discussion to the even case by using the trick of passing to
the double covering, in which one puts $ \, z_1^2 \; = \; z \, $,   with
the effect of getting formally the same result as in the  $ \, m = $
even  case.

The $ \, \omega$-foliations on  $ \, {\widehat S} \, $ extend to singular
foliations on $ \, S \, $ which determine $ \, (m+2) \, $ sectors at a
zero of $ \, \omega \, $ of order $ \, m \, $; each of these sectors has
angles equal to $ \, \pi \, $ with respect to $ \, d  s^2_\omega \,
$. Consequently,  the $ \, \omega $-metric for $ \, S \, $ is a flat
metric with a conical singularity of angle $ \, (m+2) \pi \, $ at each
zero of $ \, \omega \, $ of order $ \, m \, $.  Finally, the area of $
\,  S \, $ is equal to $ \, \| \omega \| \, $ and the $ \,
\omega$-metric  naturally induces a transverse measure on the $ \,
\omega$-foliations.

By construction, the flat metric associated to $ \, \omega \, $ admits
an atlas for $ \, {\widehat S} \, $ with values in Euclidean $ \erredue
$  in which the changes of coordinates are compositions of translations
and of the rotation of angle $ \, \pi \, $ exclusively. In fact, this
property characterizes uniquely the flat structures induced by quadratic
differentials. More precisely, given a flat metric with conical
singularities admitting such a kind of atlas on a topological surface $
\, \Sigma_g \, $, one can reconstruct a conformal structure on $ \,
\Sigma_g \, $ and a quadratic differential $ \, \omega \, $ inducing the
given flat metric. The $ \, \omega$-horizontal and the $ \,
\omega$-vertical foliations are just the pull-back on $ \, \Sigma_g \, $
of the horizontal and vertical straight lines of  $ \,
\erredue $.

A version of the Gauss-Bonnet formula gives the relation
$$
\sum_i \, m_i \; = \; - \, 2 \, \chi ( \Sigma_g ) \; = \; 4 \,  ( \, g \,
- \, 1 \, ) \qquad ,
$$
where $ \, \chi ( \Sigma_g ) \, $ is the Euler characteristic and  $ \,
\{ \, m_i \, \} \, $ are the orders of the zeros of $ \, \omega \, $. In
particular, this implies that for $ \, g =1 \, $  a quadratic differential
$ \, \omega \, $ has no zeros.

Let us now describe the deformations of the conformal  structure
according to Teichm\"uller theory.  Teichm\"uller space $ \, T_g \, $ is
the "orbifold" universal covering of the  moduli space $\, {\cal M}_g
\,$  of conformal equivalence classes of   Riemann surfaces of genus $ \,
g \, $. The elements of $\, T_g \, $ are equivalence classes of "marked"
Riemann surfaces, i.e. of homeomorphisms $ \,  \phi : \Sigma_g \to S \,
$, where $\, \Sigma_g \,$ is a fixed topological surface of genus $\, g\,
$ and $\, S \, $ is a Riemann surface. Two marked surfaces
$\, ( \, S_i , \phi_i\, ) \, $ with $\,  i=1,2 \, $ are equivalent  iff
there exists a conformal map $\, f : S_1 \to S_2\, $ such that $\, \phi_1
\circ f \circ {\phi_2}^{-1} \, $ is a automorphism of $\, \Sigma_g \,$
which is isotopic to the identity. We will  denote by $ \,  [ \, \phi :
\Sigma_g \to S \, ]\, $ the element of $\, T_g \,$ represented by the
marked surface $\, ( \, S , \phi \, ) \, $.

Let $ \, \alpha = [ \, \phi : \Sigma_g \to S \, ]
\in T_g \, $, $ \, \omega \in Q(S) \backslash \{ 0 \} \, $ and $ \,  k
\in [0,1) \, $. One can  deform the conformal structure on $ \,  S \, $
as follows: by using the coordinates $ \, \zeta \, $ defined  on $ \,
{\widehat S} \, $, set
$$
\zeta^\prime \; = \; \frac{\zeta \, + \, k \, {\overline \zeta}}{1-k}
\qquad .
$$
If $ \, \zeta = \chi + i \eta \, $, one finds $ \, \zeta^\prime =  \tau
\, \chi + i \eta \, $  where  $\,  \tau  = (1+k)/(1-k)\, $. In this way
one can define a new flat metric
$$
d  s^2_\tau \; = \; d \, \zeta^\prime \, d \, {\overline \zeta}^\prime \;
= \;  \tau^2 \, d \chi^2 \, +\,  d \eta^2
$$
on $ \, \widehat S \, $ which coincides with $\, d s^2_\omega \, $  for $
\, \tau =1 \, $. Note that any atlas for $  \left ( {\widehat S} , d
s^2_\omega \right )  $ with values in $  \left (
\erredue , \, d \chi^2 +  d \eta^2  \right )  $, with changes of
coordinates given only by combinations of translations and $ \, \pi
$-rotations, is also an atlas for $  \left ( {\widehat S} , d s^2_\tau
\right )  $ with values in $  \left ( \erredue , \, \tau^2 d \chi^2 +  d
\eta^2  \right )  $.  Therefore, $ \, d s^2_\tau \, $ induces a conformal
structure $ \, S^\omega_\tau \,
$  carrying a quadratic differential $ \, \omega_\tau \, $ such that
$$
 d   s^2_\tau  \; = \; d   s^2_{\omega_\tau } \qquad .
$$
Moreover, $ \, \omega_\tau \, $ and $ \, \omega \, $ have the same
zeros with the same orders, the horizontal and vertical foliations are
constant in $ \, \tau \, $ apart from the transverse measure: the length
of the vertical lines remains unchanged whereas the length of the
horizontal lines gets stretched by a factor $ \, \tau \, $. The line  in
Teichm\"uller space
\begin{eqnarray*}
[ \, 1 \, , \, + \infty \, ) \; & \longrightarrow & \; T_g \\
\tau \; \; \; \; \;  & \longrightarrow & \; \alpha^\omega_\tau \; =  \;
\left [ \, {\rm id } \circ
\phi \; : \; \Sigma_g \, \to \, S^\omega_\tau \, \right ]
\end{eqnarray*}
is called the Teichm\"uller ray based on $ \, ( \alpha , \omega ) \, $.
Note that the substitution of $ \, \omega \, $ by $ \, \beta \, \omega \,
$ with $ \, \beta > 0 \, $ does not modifies the Teichm\"uller ray.
Hence, the Teichm\"uller rays  are labeled by the unitary sphere $ \,
\partial Q^1 (S) \, $ in $ \, Q(S) \, $.

Now, fix $ \, \alpha = [ \, \phi : \Sigma_g \to S \, ] \in T_g \, $ and
let $ \, Q^1(S) \, $ be the unitary open ball in $ \, Q(S) \, $. Let us
define $ \, p_\alpha \, : \, Q^1(S) \to T_g
\, $ by
\[
p_\alpha (\, \omega \, ) \; = \; \left \{ \begin{array}{ll}  \alpha &
\mbox{ if $ \, \omega = 0 $}  \\
\alpha^{\widetilde \omega }_{ b (\omega )}  & \mbox{ otherwise }
\end{array}
\right. \]
where $ \, \widetilde \omega = \omega /   \| \omega \| \, $ and $ \,  b
(\omega ) = ( 1 + \| \omega
\| )/( 1- \| \omega \| ) \, $.  A fundamental result of Teichm\"uller
theory states that, for every $
\, \alpha \in T_g \, $, $ \, p_\alpha \, $ is bijective.

For every $ \, \alpha , \beta \in T_g \, $ such that $ \, \beta =
p_\alpha (\omega ) \, $, define
$$
d_T ( \, \alpha \, ,\,  \beta \, ) \; = \; \mezzo \, {\rm log} \,  \left
( \frac{1 \, + \, \| \omega
\| }{1 \, - \, \| \omega \|} \right ) \qquad .
$$
$ \, d_T \, $ is a well defined distance on $ \, T_g \, $ called the
Teichm\"uller distance;  in fact, $ \, d_T \, $ is induced by a
Finslerian metric on $ \, T_g \, $ with respect to a compatible
differential structure on $ \, T_g \, $, the Teichm\"uller rays  are
geodesic rays and each $ \, p_\alpha \, $ is a diffeomorphism.  The
cotangent bundle $ \, T^*_g \, $ is identified with
$$
T^*_g \; \approx \; \{ \, ( \, \alpha \, , \, \omega \, ) \, | \,
\alpha = [ (S, \phi )] \in T_g
\; , \; \omega \in Q(S) \, \}  \qquad .
$$
$ \, \left ( \alpha_\tau^\omega , \omega_\tau / \tau \right ) \, $  with
$ \, \tau \in (0, \infty ) \, $ is the Teichm\"uller geodetic flow on $ \,
T^*_g \, $  governed by the Lagrangian $ \,  \| \omega \|^2 / 2 \, $. Note
that we have extended each  Teichm\"uller ray  to a complete  oriented
Teichm\"uller line  by setting $ \, \tau \in \, ( 0, \infty ) \, $ in the
previous formulae.   According to this definition, a  Teichm\"uller line
based on $\, ( \alpha ,  \omega )\,
$  is just the union of two Teichm\"uller rays: one of them is based on
$\, ( \alpha ,  \omega )\, $ and the other is based on $\, ( \alpha , -
\omega )\, $.

\medskip

\noindent {\bf 3. Spacetime for a gravity system.} ~The key observation,
which allows us to associate a (2+1) gravity system to each Teichm\"uller
line, is contained in the following change of coordinates.  In \erretre
with coordinates $ \, ( u,y,\tau ) \, $, consider the upper half-plane
$ \, \Pi = \{ \, \tau > 0 \, \} \, $ with the metric of signature  $\,
(++-) \, $ given by
$$
\tau^2 \, d u^2 \, + \, d y^2 \, - \, d  \tau^2
$$
This metric is flat; indeed, under the change of coordinates
$$
x \; = \; \tau \, {\rm sh} u  \quad , \quad y \; = \; y \quad ,  \quad t
\; = \; \tau \, {\rm ch} u
\qquad ,
$$
the set $ \, \Pi \, $ goes onto the open domain $ \,  \Delta = \{ \, t >0
\, ,\,  x^2 - t^2 < 0 \, \} \, $ of the standard Minkowski space with
coordinates $ \, (x,y,t) \, $ and metric $ \, dx^2 + dy^2 - dt^2 \, $.
The constant-time hyperplanes $ \, \{ \, \tau= \tau_0 \,
\}_{\tau_0 > 0} \, $ have constant extrinsic mean curvature equal to
$\, 1/( 2  \tau_0 )\, $, so we say that they realize  a natural York
slicing of $ \, \Pi \, $. Note that the isometry group of $
\, \Pi \, $ is isomorphic with the subgroup of the Poincar\'e  group $ \,
ISO(2,1) \, $ having $ \, \Delta \, $ as invariant subset and consists of
combinations of translations parallel to the $ \, ( \tau = {\rm constant}
\, ) \, $ planes and the $ \, \pi$-rotation around the $ \, \tau \, $
axis.  Then, as $ \, \tau \, $ varies we see a one-parameter family of
flat metrics on $ \, \erredue \, $ which is formally the same occurring in
the Teichm\"uller deformation.

Consider $ \, g > 1 \, $ and the Teichm\"uller line
based on  $ \, \xi = ( \alpha , \omega ) \in T^*_g \, $,   with  $ \,
\omega \not= 0 \, $, and the homeomorphism
$$
\psi \; : \; \Sigma_g \, \times \,  ( \, 0\, , \, \infty \,  ) \;   \to
\;  M
$$
where $ \, M \, $ is a 3-manifold fibred by the surfaces   $\,
S^\omega_\tau   \equiv  \psi \, (
\, \Sigma_g \times \{ \tau \} \,  ) \,  $  with $ \, \tau \in (\, 0\, , \,
\infty \, ) \, $. Let $ \,
S^\omega_\tau \, $ be endowed with the flat metric  $ \,
ds^2_{\omega_\tau } \, $ with conical singularities.  Let us define  $ \,
{\widehat S^\omega_\tau} = S^\omega_\tau
\backslash \{ {\rm zeros ~ of~ } \omega_\tau \} \,  $ and
$ \, M^\prime = \cup_\tau  {\widehat S^\omega_\tau} \, $. One  can give $
\, M^\prime \, $ the metric
\begin{equation}
d  s^2_\xi \; = \; d s^2_{\omega_\tau } \, - \, d  \tau^2 \qquad ,
\end{equation}
which is flat and locally Minkowskian on $ \, M^\prime \, $; in fact,
it is  immediate to produce an atlas of $ \, M^\prime \, $ modeled on
the  natural York slice of $ \, \Pi \, $ so that the surfaces $ \,
{\widehat S^\omega_\tau} \, $ correspond to the $ \, \tau$-constant
hyperplanes.

Let us consider the extension of the 3-dimensional metric (1) to  the
entire manifold $\, M \, $. Since the metric (1) has vanishing
shift-vectors, the conical singularities of
$ \, ds^2_{\omega_\tau } \, $ on $ \, S^\omega_\tau \, $ survive  in
the three-dimensional context and contribute to the  three-dimensional
curvature for any $ \, \tau \, $.   We shall now prove that the
3-manifold $ \, M \, $ is flat with the exception of  the world-lines
(that we call the singular lines) associated with the zeros of the
quadratic differential  $ \, \omega \, $ ; at each zero of $ \, \omega
\,  $ of order $ \, m \, $, one has a spatial conical singularity of
angle $  \, (m+2) \pi \, $.

Consider  a tubular neighbourhood $ \, V \, $ of a singular line, $
\, V \subset M \, $; we shall denote by  $ \, V^\prime \, $ the
complement of the singular line in $ \, V \, $.  For $ \, \tau =1 \, $
suppose that, in local coordinates, one has
$ \,  \omega = z^m \, dz^2 \, $ . In a  neighbourhood of $ \, z=0 \, $,
the two-dimensional metric induced by the Teichm\"uller deformation is
given by
$$
ds^2_{\omega_\tau} \; = \; \frac{4}{(m+2)^2} \> d   \left (
\frac{z^{(m+2)/2} \, + \, k \, {\overline z}^{(m+2)/2}}{1-k} \right ) \,
d \left ( \frac{{\overline z}^{(m+2)/2} \, + \, k \,  z^{(m+2)/2}}{1-k}
\right ) \qquad .
$$
By using polar coordinates $ \, z = r \, {\rm cos} \theta \, +\,   i\,
r\,  {\rm sin} \theta \, $, one has
\begin{eqnarray*}
z^{(m+2)/2} \; &=& \; r^{(m+2)/2} \, {\rm cos} \left ( \frac{m+2}{2}
\theta \right ) \, + \, i \,  r^{(m+2)/2} \, {\rm sin} \left (
\frac{m+2}{2} \theta \right ) \\ &=& \; A(\, r \, , \, \theta \, ) \, +
\, i \, B(\, r \, , \, \theta \, ) \qquad ,
\end{eqnarray*}
and, since $ \, \tau = ( 1+k)/(1-k) \, $, one  finds
$$
ds^2_{\omega_\tau} \; = \; \frac{4}{(m+2)^2} \left \{ \, \tau^2 \,
\left ( \,  d \, A(\, r \, , \,
\theta \, ) \, \right )^2 \, + \, \left ( \,  d \, B(\, r \, , \,
\theta \, ) \, \right )^2 \, \right \} \qquad .
$$
Let us assume that the order $ \, m \, $ of the zero of  $ \, \omega \, $
is even.  We shall now prove that $ \, V^\prime \, $ is isometric with a
$(m+2)/2$-branched covering  of a locally Minkowskian manifold. Indeed,
consider  {\erretre} with coordinates $\, (x,y,t  )\, $ endowed with the
usual Minkowski metric $ \, ds^2 = dx^2 + dy^2 - dt^2 \, $ and the map
given by
\begin{eqnarray}
t \; &=& \; \tau \> {\rm ch} \left [ \, \frac{2}{m+2} \,  A(\, r \, , \,
\theta \, ) \, \right ]
\nonumber \\
x \; &=& \; \tau \> {\rm sh} \left [ \, \frac{2}{m+2} \,  A(\, r \, , \,
\theta \, ) \, \right ]  \\ y \; &=& \; \frac{2}{m+2} \> B(\, r \, , \,
\theta \, ) \qquad . \nonumber
\end{eqnarray}
The pull-back of the metric $ \, ds^2 \, $ in the coordinates  $ \, ( \,
r \, , \, \theta \, , \,
\tau \, ) \, $ is
$$
\frac{4}{(m+2)^2}\, \tau^2 \, \left ( \,  d \, A(\, r \, ,  \, \theta \,
) \, \right )^2 \, + \,
\frac{4}{(m+2)^2}\,  \left ( \,  d \, B(\, r \, , \, \theta \, )  \,
\right )^2 \, - \, d \tau^2 \; = \; ds^2_{\omega_\tau} \, - \,  d \tau^2
$$
which coincides precisely with the three-dimensional metric  $ \,
ds_\xi^2 \, $ on $ \, V^\prime \,
$. The singular line in $ \, V \, $ is mapped into the straight line  $
\, ( \, x=0 \, , \, y=0 \, )
\,  $  in {\erretre} which can be interpreted as the world-line of  a
``static particle". The map (2) corresponds to a $(m+2)/2$-fold cyclic
covering of the complement of the $ \, ( \, x=0 \, ,
\, y=0 \, ) \,  $ straight line in {\erretre}; therefore, the extension
of the metric (1) in $
\, V \, $ has a conical spatial singularity of angle $ \, (m+2) \pi \ $.
When $ \, m \, $ is odd, one can apply the same argument to the double
covering of $ \, V^\prime \, $ and one obtains the same final result;
namely, the conical singularity is of angle  $ \, (m+2) \pi \ $.

To sum up, each Teichm\"uller line is canonically associated with a
3-manifold $ \, M \, $ equipped with the metric (1) of Lorentz signature
which is locally flat and has conical spatial singularities along the
world-lines associated with the zeros of the corresponding quadratic
differential. Thus,  $ \, M \, $ can be interpreted \cite{8,9,10,11,12}
as the spacetime of a certain (2+1) gravity system containing point-like
gravitating particles moving on a two-dimensional surface of genus $ \,
g  >1\, $.  In fact, for a localized particle of mass $ \, \mu >0 \, $,
the associated  conical singularity in its rest frame has angle
\cite{11}
$$
2 \, \pi \, ( \, 1 \, - \, 4 \, G \, \mu \, ) \qquad ,
$$
where $ \, G \, $ is the three-dimensional gravitational constant.   As
noted by 't~Hooft
\cite{12}, in $(2+1)$ dimensions the sign of $ \, G \, $ is not fixed  a
priori; this property is also connected with the existence of a
Chern-Simons interpretation \cite{1} of $(2+1)$  gravity.  With positive
$ \, G \, $, the conical singularity associated  with the world-line of
a  particle of small mass $ \, \mu \, $ has angle less than $ \, 2 \pi
\,
$; whereas for negative $
\, G \, $ the conical singularity has angle greater than $ \, 2 \pi \, $.
As far as our particular gravity systems are concerned, the conical
singularities of angles $ \, (m+2) \pi \ $ in $ \, M \, $ admit a
physical interpretation in terms, for instance, of particles of mass $ \,
\mu = -m /8 G \, $  with negative $ \, G \, $. The number of particles
associated with the zeros of $ \, \omega \in Q(S)
\backslash \{ 0 \} \, $ and their masses are constrained by the
Gauss-Bonnet formula and the total mass is equal to $ \, \chi ( \Sigma_g
) / 4 G \, $.

When $ \, g =1 \, $, the same construction reproduces the nonstatic
matter-free solutions studied in \cite{2,3,4,5}. Our interpretation
based  on the Teichm\"uller flow gives also a clear explanation of the
already noted \cite{4} fact that the orbit in $ \, T_1 \, $, which is
identified with the Poincar\'e disc, is in fact geodesic.

In the remaining part of this article, we will call these universes
associated with Teichm\"uller lines (for $ \, g \geq 1 \, $) the
Teichm\"uller universes.

\medskip

\noindent {\bf 4. Asymptotic states.} ~For each Teichm\"uller universe
$ \, M \, $, which is associated to a Teichm\"uller line based on $\, (
\alpha ,  \omega )\, $, we have selected a canonical time $ \, \tau \, $
on $ \, M \, $ realizing a York slicing of its matter-free part $ \,
M^\prime \, $; the constant-time spatial surfaces have mean extrinsic
curvature $ \, 1/ (2 \tau )  \,
$ and  area $ \, \tau \| \omega \| \, $. Let us consider the asymptotic
behaviour of the universe in the  ``initial time" $ \, \tau \rightarrow 0
\, $ limit  and in the  ``final time" $ \, \tau
\rightarrow \infty \, $ limit. It seems physically interesting to note
that, for asymptotic times where time and metric do not exist, one still
has something more than just a ``topological shadow" (the genus): in
fact, metrics degenerate to measured foliations. We shall now elaborate
on this point.

As we have seen in first section, $ \, T_g \, $ is star-shaped by the
geodesic Teichm\"uller rays emanating from any fixed base point $ \,
\alpha \in T_g \, $. By adding the end-point of each ray, one gets the
so-called Teichm\"uller compactification $ \, {\overline T}_g (\alpha )
\, $. When $
\, g > 1 \, $, $ \, {\overline T}_g (\alpha )  \approx {\overline
B}^{6g-6}\, $ and this compactification actually depends on $ \, \alpha
\, $; whereas $ \, {\overline T}_1 (\alpha )
\approx {\overline B}^{2}\, $.

It is natural to identify the end-point of each ray by the associated
$\,
\omega$-vertical measured  foliation  $\, {\cal F}_v = {\cal F}_v
(\omega ) \, $. The physical interpretation of the asymptotic states,
which are defined by measured foliations,  is based on a set of
observables connected to the length of simple curves.  Indeed,  let us
denote by $ \, {\cal S} \, $ the set of isotopy classes of essential
simple curves on
$ \, \Sigma_g \, $. For each $ \, \tau \in [ 1 , \infty ) \, $ and for
each $ \, \gamma \in  {\cal S}
\, $, set
$$
\ell_\tau ( \, \gamma \, ) \; = \; \inf_{C\in \gamma } \> \ell_\tau  (\,
C \,) \qquad ,
$$
where $ \, \ell_\tau (C) \, $ is the length of the curve $ \, C \, $
with respect to the metric $
\, ds^2_\tau \, $ on $ \, \Sigma_g \, $.  Let us denote by $ \, \rho_v \,
$ the transverse measure of the foliation $ \, {\cal F}_v \, $ and
define
$$
i\, ( \, {\cal F}_v \, , \, \gamma \, ) \; = \; \inf_{C\in \gamma } \>
\rho_v ( \, C \, ) \qquad .
$$
Then, it is not hard to show that for every $ \, \gamma \in {\cal S} \,
$ one has (see \cite{9})
$$
\lim_{\tau \rightarrow \infty } \, {\ell_\tau ( \, \gamma \, )  \over
\tau} \; = \; i\, ( \, {\cal F}_v \, , \, \gamma \, ) \qquad .
$$
Each Teichm\"uller line based on  $\, ( \alpha ,  \omega )\, $
determines two end-points on the boundary of $ \, {\overline T}_g (\alpha
) \, $. The measured foliation associated with the asymptotic
$ \, \tau \rightarrow \infty \, $ final configuration is given  by  the
$\, \omega $-vertical  foliation $\, {\cal F}_v (\omega ) \, $.  Whereas
the $\, \omega $-horizontal foliation $\, {\cal F}_h (\omega ) = {\cal
F}_v (-\omega )\, $ describes the asymptotic $ \, \tau \rightarrow 0 \,
$  initial configuration. Therefore,  $\, {\cal F}_h (\omega ) \, $ and
$\, {\cal F}_v (\omega ) \, $ are interpreted as the  asymptotic states
of the Teichm\"uller universe.

When $ \, g =1 \, $, the Teichm\"uller compactification  $ \, {\overline
T}_1 (\alpha ) \, $ does not depend on $ \, \alpha \, $ and coincides
with the $ \, g =1 \, $ version of Thurston compactification. In this
case, the Teichm\"uller universes are completely determined by the
asymptotic states up to an overall rescaling.

For $ \, g > 1 \, $, the compactification $ \, {\overline T}_g (\alpha )
\, $ nontrivially depends
\cite{9} on the choice of the base point $ \, \alpha \, $. Consequently,
in order to reconstruct the spacetime of the universe, the knowledge of
the asymptotic states must be supplemented by a specific choice of the
base point $ \, \alpha \in T_g \, $.

\medskip

\noindent {\bf 5. On the phase spaces.} Several mathematical results  in
Teichm\"uller theory, see
\cite{9,15,16,17}, can be  reinterpreted  in the present (2+1) gravity
context. In this section we present a few examples.  Assume $ \, g > 1 \,
$ and $ \, G<0 \, $ as before.  Given a set of $ \, N \, $ gravitating
point-like particles moving on a surface of genus $ \, g \,
$, the type of this gravity system is, by definition,
$$
( \, g \, , \, \{ \, \alpha_i \, \} \, )
$$
where
$$
\alpha_i \; = \; 2 \, \pi \, ( \, 1 \, - \, 4 \, G \, \mu_i \, )
$$
and $ \, \mu_i > 0 \, $ is a mass for each $ \, i  =  1,2,  \cdots , N\,
$. Two basic general questions arise:
\begin{description}
\item {(1)} Determine all the couples $ \, (  g  ,  \{  \alpha_i  \}  )
\, $ which are the type of any gravity system.
\item {(2)} For each type, describe the phase space of the universes
realizing it.
\end{description}

\noindent To our knowledge, a complete answer to question (1) is not
known. As far as question (2) is concerned, it seems to be generally
accepted that the dimension of the phase space is given by
$$
12 \, g \, - \, 12 \, + \, 4 \, N \; = \;   {\rm dim}_{_{\errepic}} \,
T^*_{g,N}
$$
where $ \, T^*_{g,N} \, $ denotes the cotangent bundle of the
Teichm\"uller space for $ \, N$-punctured surfaces of genus $ \, g \, $.

These questions can be specialized in the framework of
Teichm\"uller universes; the answers in this case suggest a few general
speculations. For  Teichm\"uller universes question (1) has a complete
answer:

\begin{description}
\item {~~} ~~~$ \, (  g  ,  \{  \alpha_i  \}  ) \, $ is the type of a
Teichm\"uller universe if and only if each $ \, \alpha_i \, $ is of the
form  $ \, \alpha_i = (m_i +2) \pi \, $ with $ \, m_i \in
\Enne \, $,  $ \, m_i \geq 1 \, $, the Gauss-Bonnet relation is
satisfied and
$ \, (  g  ,  \{  \alpha_i  \}  ) \not=  (  2  ,  \{  1 , 3 \}  )$.
\end{description}

\noindent Apart from the exceptional case, we already mentioned  the
``only if" part of this statement; the ``if" part is nontrivial, see
\cite{15}. On the other hand, not every universe of this type is a
Teichm\"uller universe. For example, for any $ \, \omega \in Q(S)
\backslash \{ 0 \} \, $ consider the static universe $ \, S \times
\erreuno $ with the product metric $ \, d s^2_\omega - dt^2 \, $.

Let us now consider question (2); first of all there is a subtle
problem concerning which kind of ``isomorphism relation" one stipulates
to work with. This problem includes the assumption  on whether particles
with equal masses may be or may not be distinguished from each other.
These two possibilities reflect on the choice of the mapping class group
for the $ \, N$-punctured surfaces of genus $ \ g \, $ which must be
used in passing from the ``Teichm\"uller space" level to the ``moduli
space" level. Apart from this subtle point,  the phase space of
Teichm\"uller universes has a rather complicated structure, see
\cite{16,17}. Each admissible type $ \, (  g  ,  \{
\alpha_i = (m_i +2) \pi \}  ) \, $ splits into two augmented types  $ \,
(  g  ,  \{  \alpha_i  \} ,
\epsilon ) \, $ where $ \, \epsilon = \pm 1 \, $. $ \, T^*_{g} \, $  is
stratified according to the augmented types as follows: $ \, ( \alpha ,
\omega ) \in T^*_g \, $ belongs to the stratum of type
$ \, (  g  ,  \{  \alpha_i  \} , \epsilon ) \, $ if and only if it
determines a universe of type  $
\, (  g  ,  \{  \alpha_i  \}  ) \, $ and $ \, \epsilon =1 \, $ iff  $ \,
\omega \, $  is the square of a holomorphic  Abelian differential.  The
stratum corresponding to the type  $ \, (  g  ,  \{  \alpha_i  \} ,
\epsilon ) \, $, if nonempty,  is a submanifold of $ \, T^*_g \, $ of
real dimension
$$
4\, g \, + \,  2\, \sum_j \nu (j) \, + \, \epsilon \, - \, 3  \quad ,
$$
where, for integer $ \,  j \geq 1 \, $,  $ \, \nu (j) \, $ is the
cardinality of  $
\, \{ m_i \, | \, m_i =j \, \} \, $. This stratification is invariant
for the Teichm\"uller flow.  When at least one $ \, m_i \, $ is odd, the
stratum of type $ \, (  g  ,  \{  \alpha_i  \} , 1 ) \, $ is empty.
Whereas when all the $ \, m_i \, $ are even, the type $ \, (  g  ,  \{
\alpha_i
\}  ) \, $ correspond to two nonempty strata of different dimensions,
with the exception of $ \, (  2  ,  \{  4  \}   ) \, $ as the stratum $
\, (  2  ,  \{  4  \} , -1 ) \, $ is empty.

The maximal dimension is realized only in the stable case in which
all $
\,  m_i =1 \, $.  It should be noted that this top dimensional stratum
is a topological nontrivial open dense subset of $ \, T^*_g \, $ (in
particular, its fundamental group is nontrivial).

These facts suggest the following speculations/conjectures.
\begin{description}
\item {C1.} For a given type $ \, (  g  ,  \{  \alpha_i  \}  ) \, $
which does not correspond necessarily to a Teichm\"uller universe, the
expected  dimension
$\, {\rm dim}_{_{\errepic}}   T^*_{g,N} \, $ is realized only for a
stable situation (in a sense to be specified). For gravity systems
admitting Teichm\"uller universes, we conjecture that $\, {\rm
dim}_{_{\errepic}}   T^*_{g,N} \, $ is realized only for the type
$$
( \, g \, , \, \{ \, 1 , 1 , \cdots , 1 \, \} \, ) \qquad , \qquad N \;
= \; 4 \, g \, - 4 \qquad .
$$
\item {C2.} Even atthe ``Teichm\"uller space" level, the phase space of
a given type is topologically not trivial (in contrast with the case of
empty universes).
\item {C3.} The fact that $ \, T^*_g \, $ contains all the strata
corresponding to different Teichm\"uller universes for given $ \, g \,
$,  seems to indicate that it would be natural to look for a ``global
phase space" in which any significant change of configurations takes
place; for instance, non connected surfaces and different genera should
also be  admitted.  In fact, already in $ \, T^*_g \, $ one could
describe   decays of particles in the framework of Teichm\"uller
universes.
\end{description}

\vfill\eject

\vfill\eject


\begin{thebibliography}{99}

\bibitem{1} E.~Witten, Nucl. Phys. B 311 (1988) 46.

\bibitem{2} G.~Mess, {\it Lorentz spacetimes of constant curvature\/},
IHES preprint IHES/M/90/28.

\bibitem{3} V.~Moncrief, Journ. Math. Phys. 30 (1989) 2907; and Journ.
Math. Phys. 31 (1990) 2978.

\bibitem{4} A.~Hosoya and K.~Nakao, Class. Quant. Grav. 7 (1990) 163; and
Prog. Theor. Phys. 84 (1990) 739.

\bibitem{5} S.~Carlip, Phys. Rev. D 42 (1990) 2647.

\bibitem{6} W.~Abikoff, {\sl The Real Analytic Theory of Teichm\"uller
Space\/}, Springer-Verlag (Berlin, 1980).

\bibitem{7} O.~Lehto, {\sl Univalent Functions and Teichm\"uller Spaces},
Springer Verlag, 1987.

\bibitem{8} K.~Strebel, {\sl Quadratic Differentials}, Springer Verlag,
1984.

\bibitem{9}  S.P.~Kerckhoff,  Topology 19 N.1 (1980) 23.

\bibitem{10} A.~Staruszkiewicz, Acta Phys. Polon. 24 (1963) 734.

\bibitem{11} J.R.~Gott and M.~Alpert, Gen. Rel. Grav. 16 (1984) 243.

\bibitem{12} S.~Giddings, J.~Abbot and K.~Kuchar, Gen. Rel. Grav. 16
(1984) 751.

\bibitem{13} S.~Deser, R.~Jackiw and G.~'t~Hooft, Ann. Phys. 152 (1984)
220.

\bibitem{14} G.~'t~Hooft, Class. Quantum Grav. 9 (1992) 1335;
Class. Quantum Grav. 10 (1993) S79 and  Class. Quantum Grav. 10 (1993)
1023.

\bibitem{15}  H.~Masur and J.~Smillie, Comment Math. Helvetici 68 (1993)
289.

\bibitem{16} W.A.~Veech, Annals of Math. 124 (1986) 441.

\bibitem{17} H.~Masur and J.~Smillie, Annals of Math. 134 (1991) 455.

\end{thebibliography}
\end{document}